\documentstyle[multicol,aps,prl,epsf]{revtex}

\begin{document}

\draft

\title{
Metal-induced gap states in epitaxial organic-insulator/metal interfaces
}

\author{Manabu Kiguchi$^{1}$\cite{add}, Ryotaro Arita$^2$,  Genki Yoshikawa$^3$,
Yoshiaki Tanida$^4$, Susumu Ikeda$^1$, Shiro Entani$^1$, 
Ikuyo Nakai$^3$, Hiroshi Kondoh$^3$, Toshiaki Ohta$^3$, 
Koichiro Saiki$^{1,3}$, and Hideo Aoki$^2$
}

\address{$^1$Department of Complexity Science $\&$ 
 Engineering, 
University of Tokyo, Kashiwa, Chiba 277-8561, Japan}
\address{$^2$Department of Physics, University of Tokyo, Hongo, Tokyo 
113-0033, Japan}
\address{$^3$Department of Chemistry, University of Tokyo, Hongo, Tokyo 
113-0033, Japan}
\address{$^4$Fujitsu Laboratories Ltd., Atsugi, Kanagawa 243-0197, Japan}

\date{\today}

\maketitle

\begin{abstract}
We have shown, both experimentally and theoretically, that 
the metal-induced gap states (MIGS) can exist in epitaxially 
grown organic insulator/metal interfaces.  
The experiment 
is done for alkane/Cu(001) with an element-selective 
near edge x-ray absorption fine 
structure (NEXAFS), which exhibits a pre-peak indicative of MIGS. 
An {\it ab initio} electronic structure calculation supports the existence 
of the MIGS.  When the Cu substrate is replaced with Ni, 
an interface magnetism (spin-polarized 
organic crystal at the interface) is predicted to 
be possible with a carrier doping.
\end{abstract}

\medskip

\pacs{PACS numbers: 73.20.-r, 73.40.Ns, 71.15Mb}

\begin{multicols}{2}
\narrowtext
{\it Introduction} 
While there are mounting interests in the nature of heterointerfaces 
(solid-solid interfaces between very dissimilar materials), 
organic crystal/metal interfaces are especially intriguing 
due to their diverse possibilities not found in inorganic 
counterparts, which may facilitate in controlling and designing 
properties of interfaces. 
Electronic structures of organic/metal interfaces are 
important from a technological point 
of view as well, since the performance of devices should 
strongly depend on the electronic structure at the interface.

Recent years have in fact witnessed several interesting results 
on the electronic structure of various organic-crystal/metal interfaces. 
While it is well-known that a bulk 
C$_{60}$ crystal becomes a superconductor when doped with 
alkali metals\cite{na350}, a charge transfer was also observed for C$_{60}$ 
crystal grown on metal substrates\cite{prl71}.  Cepek {\it et al.} 
went further to suggest that 
a temperature-dependent gap structure around Fermi energy for 
a single-crystal C$_{60}$ film on Ag(001) may 
possibly be superconductivity induced by the charger transfer\cite{prl86}.  
Lukas {\it et al.} on the other hand observed 
a charged-density wave in an ordered pentacene film on Cu(110) 
and its effect on the crystal growth\cite{prl88}.

One crucial factor in such organic-insulator/inorganic-metal heterointerfaces 
is the energy level alignment at the interface, which is still far from fully 
understood. The band alignment, measured 
by photoelectron spectroscopy and Kelvin probe, has been discussed 
in terms of various effects, such as electron transfer, image effect, 
modification of the surface dipole at metal surface, 
chemical interaction and interfacial states\cite{seki}. 
However, there is no generally accepted picture for 
organic/metal interfaces, which is contrasted with 
inorganic-semiconductor/metal interfaces 
where the band alignment is known to be governed 
by the interface states called the metal-induced gap 
states (MIGS)\cite{prl52}.  So a study of the interface 
states at the organic/metal interfaces should be imperative.  

The notion of MIGS was first introduced for semiconductor/metal junctions in 
discussing the Schottky barrier\cite{prl52}. 
MIGS are roughly free-electron-like metal wave functions penetrating 
into the semiconducting side, where the penetration depth 
is inversely proportional to the band gap in 
the conventional band picture.  So a usual wisdom is that MIGS 
would be far too 
thin in insulator/metal interfaces to be relevant.  
The present authors succeeded in 
fabricating epitaxial alkali halide(insulator)/metal interfaces, which has 
enabled us to obtain unambiguous evidences especially from the 
near edge x-ray absorption fine structure (NEXAFS) 
that MIGS are in fact formed at the inorganic insulator/metal 
interfaces\cite{prl90,prb69}. 

While alkali halides are typical ionic insulators, 
interface states for {\it organic}-insulator/metal interfaces 
are a totally different issue, 
since chemical bonds of organic insulators are covalent, and it is a 
fundamental question to ask whether the formation of MIGS at 
insulator/metal interfaces are universal enough to accommodate 
organic insulators.  
Experimentally, interface states in atomically well-defined 
organic-insulator/metal interfaces has yet to be observed 
to our knowledge\cite{prl91}, despite the prime importance.  
One obvious reason for this is difficulties in fabricating atomically 
well-defined organic-insulator/metal interfaces and 
in detecting signals from the interface\cite{prl90}.  However, recent 
developments in the molecular beam epitaxy (MBE) technique have made it 
possible to prepare various types of heterointerfaces 
in case of inorganic materials\cite{mgo}, 
and novel electronic structures including MIGS 
are revealed at the interfaces\cite{prl90}. 
On the other hand, some organic films 
are begun to be epitxially grown on metal substrates in a layer-by-layer 
fashion with MBE\cite{apl84,ss515}. 

Given this background, the purpose of this Letter is 
to examine the interface states in an atomically well-defined organic 
insulator (C$_{44}$H$_{90}$) grown on Cu substrate.  
We have found, with the element-selective NEXAFS, 
a pre-peak indicative of metal-induced gap 
states.   At the same time we have performed an {\it ab initio} 
electronic structure calculation.  The theoretical result 
shows that (i) MIGS do exist at the organic insulator/metal interface, 
and (ii) when we replace Cu with Ni, in which narrow 3d bands rather 
than wide 4s bands dominate the electronic properties around the
Fermi energy, an interface magnetism (spin-polarized 
organic crystal at the interface) is predicted to be possible 
with a carrier doping.

{\it Experimental}
The experiments were performed with a UHV chamber at the soft x-ray beam 
line BL-7A in the photon factory in the Institute of Materials Structure 
Science, Tsukuba, Japan.  
We have employed an alkane (C$_{n}$H$_{2n+2}$ with $n=44$; 
tetratetracontane or TTC), 
where the properties of this series of molecules are similar 
except for a variation of the band gap with $n$.   
TTC was evaporated on Cu(001) with the substrate temperature 
of 300 K in a Knudsen cell. Real-time observation of crystallinity 
and orientation of the molecules was done with the reflection high 
energy electron diffraction (RHEED). A clear RHEED pattern is observed 
for 1 monolayer (ML) thick TTC/Cu(001).  This indicates that the TTC 
film epitaxially grows on Cu(001) with its molecular axis parallel to 
the [110] azimuth of the Cu substrates 
in a layer-by-layer fashion\cite{ss515}. Carbon $K$-edge NEXAFS spectra 
were then obtained by the partial electron yield method with a 
micro-channel plate. 

\begin{figure}
\begin{center}
\leavevmode\epsfysize=55mm \epsfbox{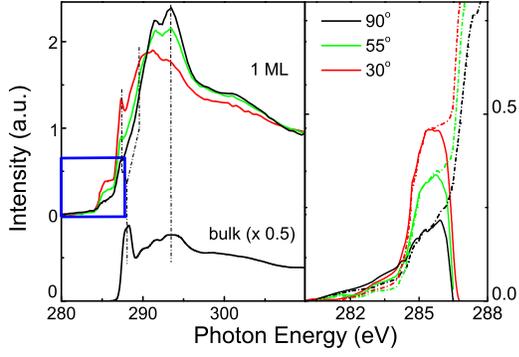}
\caption{ Experimental result for 
the C-$K$ edge NEXAFS spectra in 1 ML C$_{44}$H$_{90}$(TTC) films 
grown on Cu(001) for the x-ray incidence angle $\theta$ varied over 
30$^{\circ}$, 55$^{\circ}$, 90$^{\circ}$.  
We also show for comparison the result for a multilayer TTC (bulk)
on Cu(001) in black (which is almost $\theta$-independent 
but displayed here for $\theta =90^{\circ}$).  
All the spectra are normalized by their edge-jump. 
Right panel is a blowup of the prepeaks, 
obtained by subtracting the bulk (multilayer) spectrum.}
\label{fig1}
\end{center}
\end{figure}

\begin{figure}
\begin{center}
\leavevmode\epsfysize=50mm \epsfbox{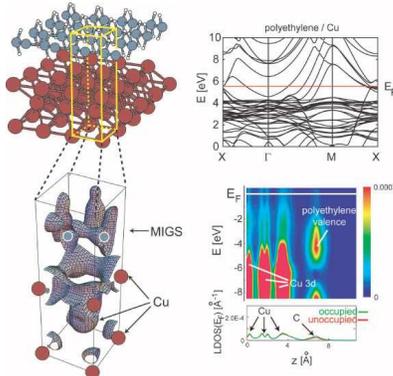}
\caption{Theoretical result for the band structure of 1 ML 
polyethlene/Cu(111). Top left inset is the atomic 
configuration considered here, where Cu is in red, C in grey, and 
H in white. Bottom left is a contour plot of 
the local density of states (LDOS), while the bottom 
right is the LDOS integrated over the $xy$ plane 
as a function of the growth direction $z$ and energy (color coded) 
along with the LDOS at $E_{F}$ versus $z$.}
\label{fig2}
\end{center}
\end{figure}

Figure~\ref{fig1} shows the NEXAFS for a 1ML 
TTC on Cu(001), as compared with the result for a bulk (multilayer) TTC film. 
A broad peak at about 293 eV can be assigned to C 1s to $\sigma^{*}$(C-C)
resonance, while the peak at 288 eV to a C 1s to $\sigma^{*}$(C-H)\cite{la17}. 
There are two points to note in the result for 1ML TTC/Cu(001). 
First, the NEXAFS result exhibits a clear polarization dependence. 
The $\sigma^{*}$(C-C)
peak, whose transition moment is parallel to the -C-C-C- plane, 
is most enhanced at the normal x-ray incidence, which confirms that 
the TTC grows on Cu(001) with its molecular axis parallel to the substrate. 
On the other hand, the $\sigma^{*}$(C-H) peak, 
whose transition moment $\perp$ -C-C-C- plane, splits into two. 
This should be because 
one-half of the hydrogen atoms in the organic molecule 
touch the substrate in the lying-down 
configuration as depicted in an inset in 
Fig.~\ref{fig2}, so that the (C-H) state splits 
due to the interaction between TTC and Cu\cite{la17}. 

More importantly, a pronounced 
pre-peak is seen to appear below the bulk edge onset, 
and this indicates a first observation of the interface states at the 
organic insulator crystal/metal interface.
Because the adsorption energy of alkanes on metal surfaces is small 
($\sim 10$ kJ/mol/CH$_{2}$ chain), 
the nature of the chemical bond at the interface should be 
a typical physisorption 
state with a weak molecule-surface interaction, as contrasted with a 
chemisorption with chemical bonds formed at the interface. 
So the pre-peak originates from the proximity to a metal 
rather than from chemical bonds, and we have here an evidence 
for the formation of MIGS at an organic insulator/metal interface.

The right panel of Fig.~\ref{fig1} is a blowup of the absorption edge 
for 1ML TTC/Cu(001). We can see that the pre-peak is greater for a 
grazing x-ray incidence than for the normal incidence. In NEXAFS, 
the electronic state whose wave function orients in the surface 
normal direction is selectively detected for grazing x-ray incidence, 
so the result indicates that the MIGS wave functions are oriented 
in the surface normal direction.
 
{\it Ab-initio calculation} 
Let us now move on to the 
first-principles (density functional 
theory) calculation for the heterointerface 
to explore the electronic structure, especially the MIGS, and 
to see how the above experimental result fits 
the theoretical picture.  
Quite recently Morikawa {\it et al.}\cite{mori} 
have performed a first principles 
calculation
for interfaces between alkane and various metals such as Cu, where 
the change in the work function and a softening of the 
CH stretching mode have been studied.  
Here we explore the local density of states, 
and also study what will happen when we replace the substrate 
with a {\it ferromagnetic} d-band metal such as Ni, for which 
the spin-density functional theory is employed.  
Our expectation for such a case is the following:  
The density of unoccupied states of Ni substrate should be much greater than 
that of Cu, because Ni is ferromagnetic (spin polarized) 
with the major part of the minority-spin 3d band 
sitting above the Fermi energy. So, naively, the intensity of MIGS for 
organic/Ni would be much stronger than that for organic/Cu. 

A penalty for doing a first-principles band calculation is, 
given the complexity of the system, we have to 
replace the alkane with a finite $n$ with 
polyethylene, which is assumed to be infinitely long.  This reduces 
the size of the unit cell, 
enabling us to perform the spin density functional study. 
We have also assumed that polyethylene chains are close-packed, 
while an experiment\cite{ss515} for TTC indicates that the real 
packing has half this density.  We adopt the exchange-correlation 
functional introduced by Perdew {\it et al.} \cite{Perdew1996}
and employ ultra-soft pseudopotentials in separable 
forms\cite{Vanderbilt90,Laasonen93}. The cut-off energy of the 
plane-wave expansion for the wave function is taken to be 42.25 Ry. 
The atomic configurations (inset of Fig.~\ref{fig2}) and the 
corresponding electronic ground states are obtained with the 
conjugate gradient scheme\cite{Yamauchi1996}.

Figure~\ref{fig2} shows the band structure along with the local density 
of states (LDOS) for polyethylene/Cu. The LDOS at $E_F$ is calculated as 
$\sum_{i}|\phi_i(x,y,z)|^2$ with the summation taken over 
the eigenstates (labeled by $i$) having energies 
$E-0.125{\rm eV}<E_i<E+0.125$eV. 
The number of sampled $k$ points is 8 with the Monkhorst-Pack method 
for the integration over the Brillouin zone\cite{Monkhorst}, 
where the bands are fitted to sinusoidal forms and the tetrahedron 
method is employed.  
We can see in the result that LDOS at $E_F$ has a peak on the carbon site, 
which indicates that MIGS are formed at the polyethylene/Cu interface.  
The wave function contour (inset of Fig.~\ref{fig2}) confirms 
this in the three-dimensional space.

\begin{figure}
\begin{center}
\leavevmode\epsfysize=60mm \epsfbox{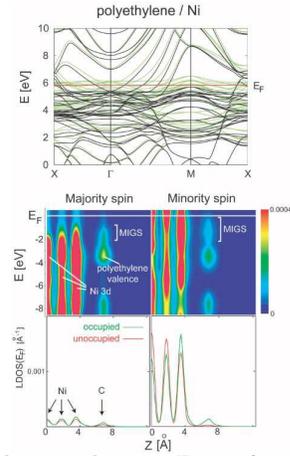}
\caption{A plot similar to Fig.\ref{fig3} for 1 ML polyethylene/Ni(111), 
where the spin density functional theory is adopted. Green (black) 
lines in the band structure represent the majority (minority) spin.}
\label{fig3}
\end{center}
\end{figure}

\begin{figure}
\begin{center}
\leavevmode\epsfysize=45mm \epsfbox{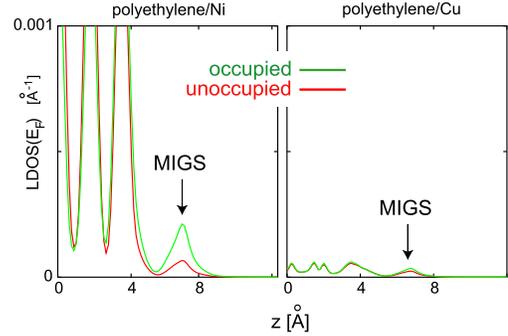}
\caption{Theoretical results for the local density of occupied and 
unoccupied states at $E_{F}$ are compared for 
polyethylene/Ni and polyethylene/Cu.} 
\label{fig4}
\end{center}
\end{figure}

{\it Ni substrate}  
If we now come to polyethylene/Ni in Fig.~\ref{fig3}, 
we have to look at the majority- and minority-spin components separately, 
since the Ni substrate is spin polarized. We can see that the 
local density 
of states around $E_F$ significantly differs between 
the majority and minority spins, including those at the 
carbon sites.  In the energy-resolved LDOS (color-coded panel 
in Fig.~\ref{fig3}) this is seen to come from 
a difference in the position of MIGS in the occupied $(E<E_F$) side.  
To be more precise, there is a finite exchange 
splitting for the MIGS in polyethylene/Ni interface, although the 
splitting is smaller than the splitting within the Ni substrate.  
However, both the majority- and minority-spin MIGS lie below 
the $E_F$ (with the latter lying just below $E_F$). 
This implies that the organic crystal, although lying on 
a ferromagnetic substrate, is not spin-polarized.
Figure \ref{fig4} compares for Ni and Cu substrates 
the total (i.e., sum of the majority- and 
minority-spin) LDOS (which is relevant to the NEXAFS). 
In accord with the above, the density of 
unoccupied MIGS states is similar between polyethylene/Ni and 
polyethylene/Cu, while the density of occupied MIGS states 
differs between them because the 3d band resides just below $E_F$ for Ni.  
The theoretical prediction agrees with our preliminary NEXAFS result, 
a probe for unoccupied states, 
for octane(C$_{8}$H$_{18}$) on Ni(111) and Cu(111) substrates. 
The intensity of pre-peak is similar between the two systems, 
with the intensity normalized by the edge jump 
being 0.27 for octane/Ni(111) against 0.30 for octane/Cu(111).

\begin{figure}
\begin{center}
\leavevmode\epsfysize=45mm \epsfbox{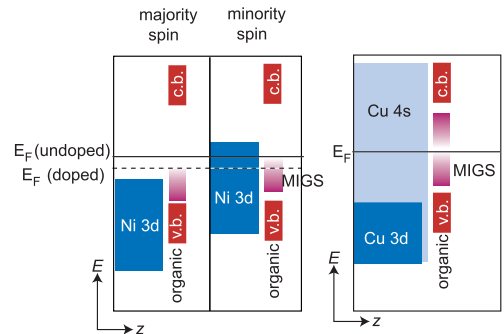}
\caption{A schematic energy diagram for MIGS in organic/Ni 
and in organic/Cu.}
\label{fig5}
\end{center}
\end{figure}

We can summarize the band scheme in Fig.~\ref{fig5}, which schematically 
depicts the energy regions for MIGS in the 
organic/Cu and organic/Ni interfaces. 
The MIGS band for the majority spin in polyethylene/Ni lies 
below $E_F$, so the density of states 
in the unoccupied side is small, while the density of occupied states is large. 
By contrast, $E_F$ runs right through the MIGS band when the substrate is Cu, 
but the density of MIGS is relatively low due to a low 
density of states of the Cu 4s band.  This results in little 
difference between organic/Ni and organic/Cu as far as the density 
of {\it unoccupied} MIGS is concerned, so an experimental method 
that detects occupied states, such as resonant photoemission or 
x-ray emission spectroscopy, will probe the difference. 

While the spin-{\it un}polarized MIGS on a ferromagnetic substrate 
is a bit of a disappoinment, the above picture 
naturally leads us to the following theoretical proposal. 
In the energy diagram in Fig.\ref{fig5}, 
we can make the MIGS spin-{\it polarized} 
if we can introduce carriers into MIGS by, e.g., chemical doping.  
Namely, 
while normally the majority and minority spins are (almost) fully occupied 
in organic/Ni, the doped carriers will accommodated in the minority-spin band 
as indicated by a dashed line in Fig.\ref{fig5}, 
so that we should end up with a polarized organic molecules.


The calculation was performed with TAPP
(Tokyo ab-initio program package), for which RA received
technical advices from Y. Suwa.  Numerical calculations
were performed on SR8000 in ISSP, University of Tokyo.
This work was supported in part by a creative scientific research 
project No. 14GS0207 and a special coordination fund from the Japanese 
Ministry of Education.

\end{multicols}
\end{document}